\documentclass[journal,twoside,web]{ieeecolor}
\usepackage{tmi}
\usepackage{cite}
\usepackage{amsmath,amssymb,amsfonts}
\usepackage{algorithmic}
\usepackage{graphicx}
\usepackage{textcomp}
\usepackage{graphics}
\usepackage{color, colortbl}
\usepackage{upgreek}
\usepackage{optidef}
\usepackage{siunitx}
\graphicspath{{./figures/}}
\usepackage{xr}
\usepackage{threeparttable}

\usepackage{etoolbox}
\makeatletter
\@ifundefined{color@begingroup}%
  {\let\color@begingroup\relax
   \let\color@endgroup\relax}{}%
\def\fix@ieeecolor@hbox#1{%
  \hbox{\color@begingroup#1\color@endgroup}}
\patchcmd\@makecaption{\hbox}{\fix@ieeecolor@hbox}{}{\FAILED}
\patchcmd\@makecaption{\hbox}{\fix@ieeecolor@hbox}{}{\FAILED}

\def\BibTeX{{\rm B\kern-.05em{\sc i\kern-.025em b}\kern-.08em
    T\kern-.1667em\lower.7ex\hbox{E}\kern-.125emX}}
\begin{document}
\title{Cram\'er-Rao Bound Optimized Subspace Reconstruction in Quantitative MRI}
\author{Andrew~Mao,
        Sebastian~Flassbeck,
        Cem~Gultekin,
        Jakob~Assl\"ander
    \thanks{This work was supported by NIH grants F30~AG077794, T32~GM136573, and performed under the rubric of the Center for Advanced Imaging Innovation and Research (CAI2R), an NIBIB National Center for Biomedical Imaging and Bioengineering (NIH P41~EB017183). (Corresponding author: Andrew Mao). \\
    Andrew Mao, Sebastian Flassbeck, Jakob Assl\"ander, are with the Center for Biomedical Imaging, NYU School of Medicine, New York, NY 10016. (e-mails: andrew.mao@nyumc.org, sebastian.flassbeck@nyumc.org, jakob.asslaender@nyumc.org).\\
    Cem Gultekin is with the Courant Institute of Mathematical Sciences, New York University, New York, NY 10012. (e-mail: cg3306@nyu.edu)}
}
\maketitle

\begin{abstract}
We extend the traditional framework for estimating subspace bases that maximize the preserved signal energy to additionally preserve the Cram\'er-Rao bound (CRB) of the biophysical parameters and, ultimately, improve accuracy and precision in the quantitative maps. To this end, we introduce an \textit{approximate compressed CRB} based on orthogonalized versions of the signal's derivatives with respect to the model parameters. This approximation permits singular value decomposition (SVD)-based minimization of both the CRB and signal losses during compression. Compared to the traditional SVD approach, the proposed method better preserves the CRB across all biophysical parameters with negligible cost to the preserved signal energy, leading to reduced bias and variance of the parameter estimates in simulation. In vivo, improved accuracy and precision are observed in two quantitative neuroimaging applications, permitting the use of smaller basis sizes in subspace reconstruction and offering significant computational savings.
\end{abstract}

\begin{IEEEkeywords}
quantitative MRI, magnetic resonance fingerprinting, magnetization transfer, Cram\'er-Rao bound, subspace reconstruction, singular value decomposition.
\end{IEEEkeywords}

\section{Introduction}\label{sec:intro}
\IEEEPARstart{Q}{uantitative} MRI (qMRI) involves a series of measurements, e.g., images with different contrasts, that encode information about the tissue's biophysical parameters, such as the tissue's relaxation times. Steady-state sequences provide an efficient way to sequentially sample the entire k-space of each contrast\cite{Deoni2003}. 
By comparison, transient-state sequences, e.g., traditional parameter mapping techniques such as multi-echo spin echo\cite{Hennig1988} or Look-Locker\cite{Look1970}, sample only small parts of each contrast's k-space during an RF-pulse train and then repeat the pulse train to fill each k-space.
Recent work aims to reduce the number of k-space samples per contrast in order to reduce the overall scan time by incorporating advanced image reconstruction techniques. These approaches include MR fingerprinting (MRF)\cite{Ma2013}, MR-multitasking\cite{Christodoulou2018}, echo-planar time-resolved imaging\cite{Wang2019epti,Wang2022}, MR-STAT\cite{Sbrizzi2018}, and hybrid-state free precession \cite{Asslander-hsCommPhysics}. 

A common approach for reconstructing such undersampled dynamic data is to use different sub-sampling patterns for each contrast and utilize redundant information between contrasts, e.g., with low-rank subspace reconstruction techniques\cite{Christodoulou2018,Liang2007,Huang2012,McGivney2014,Tamir2017,Asslander2018,Zhao2015,Zhao2018}. This approach assumes that the different contrasts lie in some low-dimensional subspace and reconstructs \textit{coefficient images} corresponding to the basis functions. This subspace is usually precomputed by taking the singular value decomposition (SVD) of a simulated dictionary of signals over the expected range of tissue parameters, e.g., using the Bloch equations or the extended phase graph method\cite{Hennig1988,Hennig1991,Weigel2015,Malik2018}.

In qMRI, high-quality image reconstruction is desirable, but not in itself as---outside of nonlinear inversion approaches\cite{Sbrizzi2018,Haldar2009,Zhao2016,Wang2018mbir,Scholand2023}---it is only an intermediate step to obtaining quantitative parameter maps\cite{Asslander2020,Shafieizargar2023}.
The most common approach is to fit the biophysical model directly to the coefficient images, e.g., with dictionary matching\cite{Ma2013,McGivney2014,Jiang2015a}, non-linear least squares, kernel regression\cite{Nataraj2018}, or neural-network-based\cite{Virtue2018,Cohen2018a,Zhang2022} fitting. In this paradigm, the subspace coefficients of each voxel are the measurements used for parameter estimation. Consequently, the variance of an unbiased estimator is bounded by the Cram\'er-Rao lower bound\cite{Cramer} (CRB) of the measured coefficients rather than the uncompressed data (e.g., a time series of images). In this article, we analyze the conservation of the CRB during the projection of the signal onto the subspace and present a strategy for optimizing the subspace to mitigate the associated increase in CRB.

CRB analysis has a rich history in MRI
where it has often been used to design the optimal experimental parameters, such as the flip angle train or the undersampling pattern\cite{Jones1997,Teixeira2018,Zhao2019,Asslander-hsMRM,Liu2017,Lee2019,Nataraj2017a,Haldar2019,Whitaker2020}. In the context of subspace reconstruction, recent work analyzed the g-factor for a CRB-optimized pulse sequence and fixed basis to identify the optimal subspace size for reconstruction\cite{Wang2023}. In this work, we optimize the basis itself, hypothesizing that, while minimum variance unbiased estimation is generally unachievable in qMRI, improved CRB preservation leads to improved conditioning of the parameter estimation process. While reduced CRBs might be expected simply to improve parameter precision, we show that, in some situations, it can also lead to improved parameter accuracy. This is consistent with viewing the CRB as a measure of the local curvature of the log-likelihood function over parameter space\cite{Asslander2020}, where higher curvature can reduce bias\cite{Jelescu2016}. In joint estimation problems, lower CRBs for difficult-to-estimate parameters can also decrease the bias sensitivity for easier-to-estimate parameters due to reduced information coupling\cite{Hero1992}.

In this work, we propose a method for optimizing linear bases to preserve the CRB of the biophysical parameters in addition to the signal energy by leveraging the CRB's geometric interpretation\cite{Scharf1993}. We incorporate orthogonalized versions of the signal's derivatives into a basis optimization scheme as an approximation for the compressed domain CRB, which we analyze in detail. We demonstrate our approach's ability to improve the accuracy and precision of the parameter maps with smaller subspace sizes in silico and in vivo in two qMRI neuroimaging applications: a) a two-pool qMT model\cite{Henkelman1993,Helms2009,Asslander2021} using a hybrid-state sequence\cite{Asslander2023}, and b) single-compartment $T_1$ and $T_2$ mapping using the MRF inversion recovery fast imaging with steady-state precession (MRF-FISP) technique\cite{Jiang2015a}. This is an expansion of our previous work\cite{Mao2023}, where we considered only the qMT model and did not analyze the approximation made to the compressed CRB.

\section{Theory}\label{sec:theory}
\subsection{Problem Description}\label{subsec:probdesc}

In this article, we adopt the following linear subspace model, which does not consider the MR imaging process:
\begin{equation*}
    \mathbf{c}=\mathbf{U}'\mathbf{s}+\boldsymbol\epsilon,
\end{equation*}
where $\mathbf{s} \in \mathbb{C}^{N_T}$ is the measured signal with $N_T$ time points or data frames, $\mathbf{U} \in \mathbb{C}^{N_T \times N_c}$ is the subspace of rank $N_c$ where $N_c \ll N_T$, $'$ is the conjugate transpose, $\mathbf{c} \in \mathbb{C}^{N_c}$
is the coefficient vector representing the contribution of each basis element to the spin evolution,
and $\boldsymbol\epsilon \sim \mathcal{N}(\mathbf{0}, \sigma^2 \mathbf{I})$ is white Gaussian noise ($\mathbf{I}$ is the identity matrix). Note that $\mathrm{Cov}(\mathbf{U}'\boldsymbol\epsilon)=\mathbf{U}' \sigma^2 \mathbf{I} \mathbf{U} = \sigma^2 \mathbf{I}=\mathrm{Cov}(\boldsymbol\epsilon)$ if $\mathbf{U}'\mathbf{U}=\mathbf{I}$, i.e., the low-rank operator does not change the noise covariance if $\mathbf{U}$ encompasses an orthonormal basis.

As the coefficients $\mathbf{c}$ are the measurements received by the parameter estimator, we use this model to generate an expression for the compressed CRB in Section~\ref{subsec:TheoryCompressedCRB}.

\subsection{Traditional SVD Basis}
The subspace for a qMRI experiment can be estimated \textit{a priori}. Using an appropriate signal model, $N_s$ simulated signal evolutions (or fingerprints) for the expected range of parameters can be stacked column-wise to form a dictionary matrix $\mathbf{S} \in \mathbb{C}^{N_T \times N_s}$. Then, $\mathbf{U}$ is the solution to the following sample principal components analysis problem:
\begin{mini}
    {\mathbf{U}}{\big\| \mathbf{S}-\mathbf{UU}'\mathbf{S}\big \|^2_F}{}{}
    \addConstraint{\mathbf{U}'\mathbf{U}=\mathbf{I}}{}{}
    \label{eq:pca}
\end{mini}
where $\mathbf{UU}'$ represents the projection onto $\langle \mathbf{U} \rangle$ (angular brackets $\langle \cdot \rangle$ denoting the column span) and $\|\cdot\|_F$ denotes the Frobenius norm which captures the signal energy loss between the compressed and original fingerprints.
Eq.~\eqref{eq:pca} is efficiently solved by taking the first $N_c$ left singular vectors of the SVD of $\mathbf{S}$ to form $\mathbf{U}$ \cite{Vidal_2016}, which we refer to as the ``traditional SVD'' approach.


\subsection{The Cram\'er-Rao Bound and its Geometric Interpretation}
For an unbiased estimator, the CRB is the minimum variance of a parameter's estimate. For Gaussian noise, each parameter's CRB is proportional to the corresponding diagonal entry of the inverse Fisher Information Matrix $\mathbf{F}^{-1}=\left(\mathbf{J}'\mathbf{J}\right)^{-1}$, where $\mathbf{J}$ is the Jacobian matrix whose columns are the signal's derivatives with respect to the parameters $\boldsymbol\theta$, i.e., $\mathbf{j}_i \triangleq \partial \mathbf{s}/\partial \boldsymbol\theta_i$ where we consider, for the moment, the uncompressed signal $\mathbf{s}$ as the measurements.

\begin{figure}[tbp]
    \centering
    \includegraphics[width=0.45\textwidth]{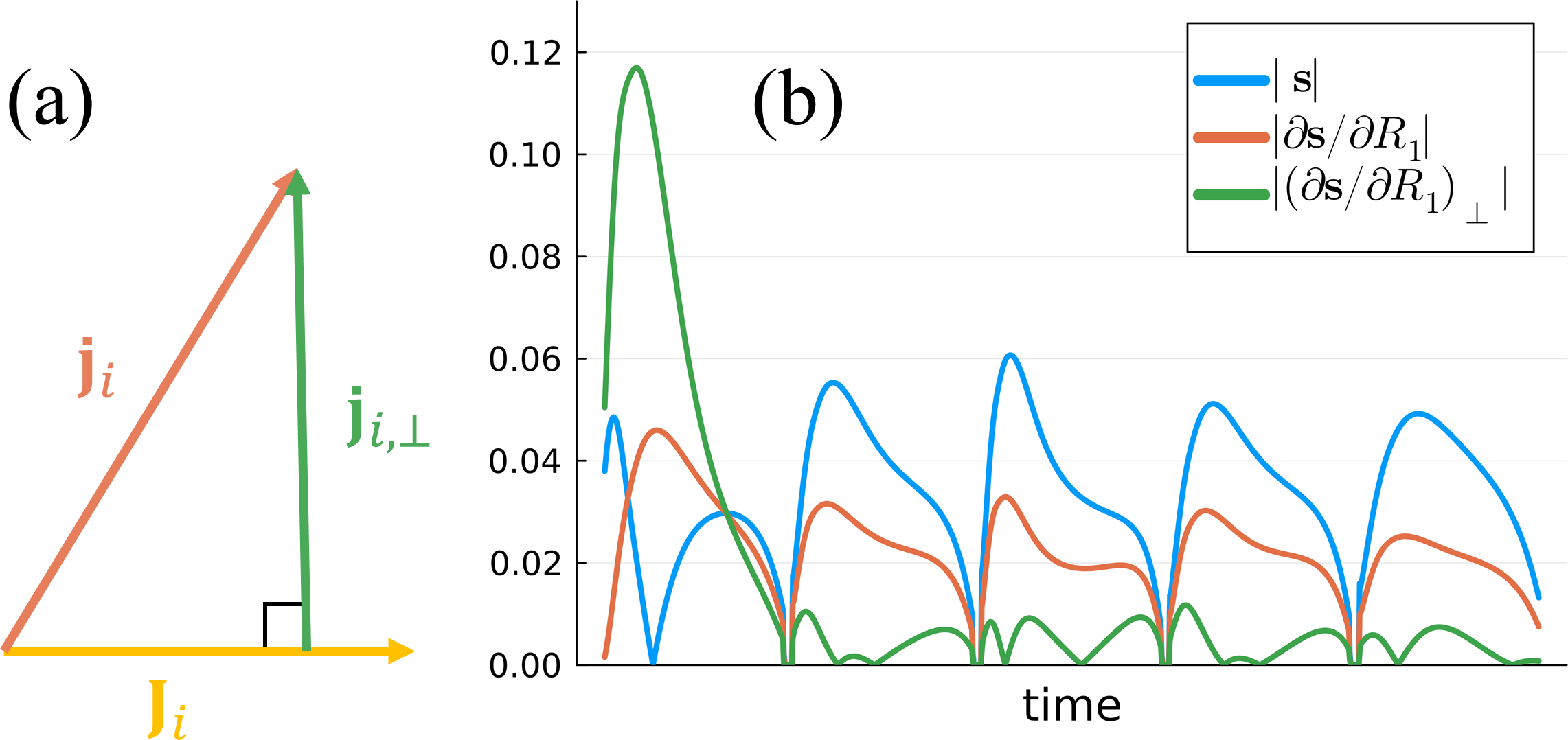}
    \caption{(a) Distinguishing one model parameter from another depends on the components of its signal derivative, $\mathbf{j}_i$, that are orthogonal to the span of all other signal derivatives, $\mathbf{J}_i$ \cite{Scharf1993}. (b) Depiction of a representative signal $\mathbf{s}$ for the inversion recovery MRF-FISP sequence \cite{Jiang2015a} (assuming the scaling $M_0=1$), its derivative with respect to the longitudinal relaxation rate $\partial\mathbf{s}/\partial R_1$, and the corresponding orthogonalized derivative $(\partial\mathbf{s}/\partial R_1)_\perp$, where the components parallel to the derivatives with respect to all other model parameters were removed (i.e., $M_0$ and $R_2$). Note we plot the absolute value and $(\partial\mathbf{s}/\partial R_1)_\perp$ is scaled up to have a unit length, which emphasizes that the first segment is most important for encoding $R_1$.}
    \label{fig:gcrb}
\end{figure}

For a one-parameter model, the CRB is simply proportional to the inverse $\ell_2$-norm of the derivative: $$B(\theta) = \frac{\sigma^2}{\mathbf{j}' \mathbf{j}}.$$
As described in Ref.~\cite{Scharf1993}, this simple notation can be translated to the multi-parametric case by considering the angle between the different signal derivatives.
Specifically, the CRB in the multi-parametric case is inversely proportional to the $\ell_2$-norm of the derivative w.r.t. $\boldsymbol\theta_i$ after removing all components parallel to the linear space spanned by the derivatives w.r.t. all other model parameters, which we denote using the matrix $\mathbf{J}_i$. 
Concretely, the \textit{uncompressed} Cram\'er-Rao bound $B_u(\boldsymbol\theta_i)$ of the parameter $\boldsymbol\theta_i$ can be written as
\begin{equation}
    B_u(\boldsymbol\theta_i) = \frac{\sigma^2}{\mathbf{j}'_i \left(\mathbf{I}-\mathbf{P}_{\mathbf{J}_i}\right)' \left(\mathbf{I}-\mathbf{P}_{\mathbf{J}_i}\right)\mathbf{j}_i} = \frac{\sigma^2}{\mathbf{j}'_{i,\perp} \mathbf{j}_{i,\perp}}
    \label{eq:tcrb}
\end{equation}
\begin{equation}
    \mathbf{j}_{i,\perp} \triangleq \left(\mathbf{I}-\mathbf{P}_{\mathbf{J}_i}\right)\mathbf{j}_i
\end{equation}
where $\mathbf{P}_{\mathbf{J}_i} \triangleq \mathbf{J}_i(\mathbf{J}_i'\mathbf{J}_i)^{-1}\mathbf{J}_i'$ denotes the projection matrix onto $\langle \mathbf{J}_i \rangle$.
To highlight the similarity to the CRB of the single-parameter model, we defined the \textit{orthogonalized derivative} $\mathbf{j}_{i,\perp}$ as the projection of the derivative $\mathbf{j}_i$ onto the space orthogonal to $\langle \mathbf{J}_i \rangle$, as shown in Fig.~\ref{fig:gcrb}. For clarity we denoted the projection matrix twice, noting that they are symmetric and idempotent, i.e., 
$\left(\mathbf{I}-\mathbf{P}_{\mathbf{J}_i}\right)' \left(\mathbf{I}-\mathbf{P}_{\mathbf{J}_i}\right) = \mathbf{I}-\mathbf{P}_{\mathbf{J}_i}$. 
Eq.~\eqref{eq:tcrb} enables the calculation of a parameter's CRB from the $\ell_2$-norm of its orthogonalized derivative alone---without requiring a matrix inverse---which helps formulate a minimization procedure. It also reveals that the orientations and norms of the spaces $\langle \mathbf{j}_i \rangle$ and $\langle \mathbf{J}_i \rangle$ are what determine the CRB. Increasing the dimension of $\langle \mathbf{J}_i \rangle$, e.g. by adding additional model parameters, usually decreases $\|\mathbf{j}_{i,\perp}\|_2$ and consequently increases the CRB. Decreasing the dimensionality, e.g. by fixing certain parameters, usually has the opposite effect.

\subsection{The Cram\'er-Rao Bound for Subspace Coefficients as Measurements} \label{subsec:TheoryCompressedCRB}
In this section, we modify Eq.~\eqref{eq:tcrb} to incorporate subspace modeling. In this case, we consider the subspace coefficients $\mathbf{c}$ as the measurements used for parameter estimation. The modification consists of simply multiplying the signal derivatives with $\mathbf{U}'$, yielding the \textit{exact compressed CRB} ($B_{ec}$)
\begin{equation}
    \begin{split}
    B_{ec}(\boldsymbol\theta_i) &= \frac{\sigma^2}{\mathbf{j}'_i \mathbf{U} \left(\mathbf{I}-\mathbf{P}_{\mathbf{U}'\mathbf{J}_i}\right)' \left(\mathbf{I}-\mathbf{P}_{\mathbf{U}'\mathbf{J}_i}\right) \mathbf{U}' \mathbf{j}_i} \\
    &= \frac{\sigma^2}{\mathbf{j}'_i \left(\mathbf{I}-\mathbf{P}_{\mathbf{UU}'\mathbf{J}_i}\right)' \mathbf{UU}' \left(\mathbf{I}-\mathbf{P}_{\mathbf{UU}'\mathbf{J}_i}\right) \mathbf{j}_i},
    \end{split}
    \label{eq:ccrb}
\end{equation}
where the latter equality follows from definition of the projection above. 
Eq.~\eqref{eq:ccrb} shows that the CRB is increased relative to Eq.~\eqref{eq:tcrb} if the projection onto $\langle \mathbf{U} \rangle$ reduces the angle between $\langle \mathbf{j}_i \rangle$ and $\langle \mathbf{J}_i \rangle$. 
However, the CRB remains unchanged if $\langle \mathbf{J} \rangle \in \langle \mathbf{U} \rangle$, i.e., if the basis captures the span of the entire Jacobian matrix.
While this condition is fulfilled if the signal $\mathbf{S} \in \langle\mathbf{U}\rangle$, this is generally unachievable when $N_c\ll N_T$, meaning there are also no guarantees surrounding the preservation of the Jacobian. 
In fact, maximizing the preserved signal energy in Eq.~\eqref{eq:pca} equates to optimizing only for the signal's derivative with respect to the scaling $M_0$ without accounting for any other derivative and their geometric relations.

\subsection{Cram\'er-Rao Bound Optimized Bases}
\label{subsec:crbbases}
To both represent the signal evolution accurately and preserve the CRB of the biophysical parameters, we propose to optimize a linear basis $\hat{\mathbf{U}}$ by incorporating both objectives in a composite cost function, given by
\begin{mini}
    {\mathbf{U}}{\left(1-\lambda\right)\big\|\mathbf{S}-\mathbf{UU}'\mathbf{S}\big\|^2_F + \lambda\big\|\mathbf{J}_\perp-\mathbf{UU}'\mathbf{J}_\perp\big\|^2_F}{}{}
    \addConstraint{\mathbf{U}'\mathbf{U}=\mathbf{I}}{}{}
    \label{eq:objective}
\end{mini}
where $\mathbf{J}_\perp \in \mathbb{C}^{N_T \times N_s N_p}$ is the matrix whose columns are the $N_p$ orthogonalized derivatives of interest for each dictionary fingerprint, each normalized to have unit energy. 
This formulation permits optimizing for only a subset of the model parameters (i.e., the \textit{parameters of interest} for any given application) while still considering a fit of the full model in computing the CRB.
Normalization of each orthogonalized derivative allows us to measure the proportion of the CRB lost in compression and weights each parameter in the cost equally, where it would otherwise be dominated by the most difficult-to-estimate parameters. While we have not done so here, we note that it is still possible to incorporate weightings for the different parameters into Eq.~\eqref{eq:objective}. $\lambda \in \left[0,1\right]$ controls the convex combination of the signal energy and CRB losses in the overall cost. Defining $\lambda$ in this way enables us to probe the behavior of the bases over a fixed range of $\lambda$ values across different pulse sequences, which may have different signal or derivative amplitudes. Setting $\lambda=0$ reduces Eq.~\eqref{eq:objective} to the traditional SVD approach in Eq.~\eqref{eq:pca}.

The second term in Eq.~\eqref{eq:objective} captures the \textit{approximate CRB loss} ($\Delta B_{ac}$) and can be rewritten as
\begin{equation}
    \begin{split}
    \big\| & \mathbf{J}_\perp-\mathbf{UU}'\mathbf{J}_\perp\big\|^2_F \triangleq \Delta B_{ac} \\
    &= \sum_{i=1}^{N_sN_p} \left( 1 - \frac{\mathbf{j}'_i \left(\mathbf{I}-\mathbf{P}_{\mathbf{J}_i}\right)'\mathbf{UU}' \left(\mathbf{I}-\mathbf{P}_{\mathbf{J}_i}\right)\mathbf{j}_i}{\mathbf{j}'_i\left(\mathbf{I}-\mathbf{P}_{\mathbf{J}_i}\right)' \left(\mathbf{I}-\mathbf{P}_{\mathbf{J}_i}\right)\mathbf{j}_i} \right) \\
    &= \sum_{i=1}^{N_sN_p} \left( 1 - \frac{B_u(\boldsymbol\theta_i)}{B_{ac}(\boldsymbol\theta_i)} \right),
    \end{split}
    \label{eq:acrbloss}
\end{equation}
where the right-hand term is the ratio of the uncompressed CRB to an \textit{approximate compressed CRB} ($B_{ac}$), defined as
\begin{equation}
    B_{ac}(\boldsymbol\theta_i) \triangleq \frac{\sigma^2}{\mathbf{j}'_i \left(\mathbf{I}-\mathbf{P}_{\mathbf{J}_i}\right)' \mathbf{UU}' \left(\mathbf{I}-\mathbf{P}_{\mathbf{J}_i}\right)\mathbf{j}_i} = \frac{\sigma^2}{\mathbf{j}'_{i,\perp} \mathbf{UU}' \mathbf{j}_{i,\perp}}.
    \label{eq:acrb}
\end{equation}
Comparing Eq.~\eqref{eq:ccrb} and Eq.~\eqref{eq:acrb} shows that the approximation lies in the projection onto $\langle\mathbf{J}_i\rangle$ instead of $\langle\mathbf{UU}'\mathbf{J}_i\rangle$. The \textit{exact compressed CRB loss} ($\Delta B_{ec}$) analogous to Eq.~\eqref{eq:acrbloss} is
\begin{equation}
    \Delta B_{ec} \triangleq \sum_{i=1}^{N_sN_p} \left( 1 - \frac{B_u(\boldsymbol\theta_i)}{B_{ec}(\boldsymbol\theta_i)} \right).
    \label{eq:ccrbloss}
\end{equation}
In general, the following relation holds:
\begin{equation}
    B_u(\boldsymbol\theta_i) \leq B_{ac}(\boldsymbol\theta_i) \leq B_{ec}(\boldsymbol\theta_i),
    \label{eq:inequality}
\end{equation}
where the latter inequality follows directly from
\begin{equation}
    \big\|\mathbf{UU}'\mathbf{P}_{\mathbf{J}_i}\mathbf{j}_i\big\|_2 \leq \big\|\mathbf{UU}'\mathbf{P}_{\mathbf{UU}'\mathbf{J}_i}\mathbf{j}_i\big\|_2,
    \label{eq:inequality_norm}
\end{equation}
because the projection onto $\langle\mathbf{U}\rangle$, a subspace of $\mathbb{C}^{N_T}$, can only decrease the angle between $\langle\mathbf{J}_i\rangle$ and $\mathbf{j}_i$. If the angle is zero after compression,
i.e. $ \langle \mathbf{UU}' \mathbf{J}_i \rangle = \langle \mathbf{UU}' \mathbf{j}_i \rangle$,
$B_{ac}$ would be finite while $B_{ec}$ is infinite. 
The following relationship also follows directly from Eq.~\eqref{eq:inequality}:
\begin{equation}
    \Delta B_{ac} \leq \Delta B_{ec},
    \label{eq:inequality_loss}
\end{equation}
meaning that minimizing the approximate CRB loss $\Delta B_{ac}$ does not in general guarantee minimization of the exact CRB loss $\Delta B_{ec}$.
However, this is less likely to occur for sufficiently large values of $\lambda$ as $\langle \mathbf{J} \rangle$ is increasingly well-preserved within the CRB-SVD basis, and hence $B_{ac}\approx B_{ec}$.

While we would ideally like to directly minimize $\Delta B_{ec}$ in Eq.~\eqref{eq:ccrbloss}, we introduce $\Delta B_{ac}$ to utilize the established and numerically preferable SVD framework to optimize the basis functions. We first compute the orthogonalized derivatives $\left\{ \mathbf{j}_{i,\perp} \right\}_{i=1}^{N_sN_p}$, which do not depend on $\mathbf{U}$. Thereafter, Eq.~\eqref{eq:objective} can be solved by simply combining the two loss terms and performing an SVD on the horizontally concatenated matrix
\begin{equation}
    \begin{gathered}
    \mathbf{D} \triangleq \begin{bmatrix} \left(1-\lambda\right)\mathbf{S} & \lambda \mathbf{J}_\perp \end{bmatrix} \\
    \mathbf{D} = \tilde{\mathbf{U}} \tilde{\mathbf{\Sigma}} \tilde{\mathbf{V}}' \\
    \hat{\mathbf{U}} = \mathbf{\tilde{U}}_{N_c}
    \end{gathered}
\label{eq:concatmatrix} 
\end{equation}
where $\mathbf{\tilde{U}}_{N_c}$ denotes the first $N_c$ columns of $\mathbf{\tilde{U}}$. Eq.~\eqref{eq:concatmatrix} shows that the primary difference between the CRB-SVD and the traditional SVD approach is the explicit inclusion of the model parameter's orthogonalized signal derivatives---rather than only the signal (i.e., $\partial\mathbf{s}/\partial M_0$)---in the subspace estimation process. This helps preserve the orientations of the subspaces $\langle \mathbf{j}_i \rangle$ and $\langle \mathbf{J}_i \rangle~\forall i$, and therefore the CRB, at some cost to the preserved signal energy.

The SVD of $\mathbf{D}$ naturally yields optimal (within the approximation), orthogonal bases which need only be computed once per value of $\lambda$ and can be retrospectively truncated according to the choice of $N_c$. Note that because the column dimension of $\mathbf{D}$ scales linearly with $N_sN_p$, assuming $N_T < N_sN_p$, the memory complexity $\mathcal{O}\left(N_TN_sN_p\right)$ and compute time complexity $\mathcal{O}\left(N_T^2N_sN_p\right)$ of the CRB-SVD scale linearly with $N_sN_p$\cite{Li2019}.


\section{Methods}\label{sec:methods}

\subsection{Pulse Sequences}
\label{subsec:sequences}

The first qMRI application we consider is the hybrid-state sequence described in Ref. \cite{Asslander2023} designed to extract the parameters of a 2-pool qMT model\cite{Henkelman1993,Helms2009,Asslander2021}, i.e. a complex-valued scaling $M_0$, the fractional semi-solid spin-pool size $m_0^s$, the free spin-pool relaxation rates $R_1^f,R_2^f$, the exchange rate $R_\text{x}$, the semi-solid spin-pool relaxation rates/times $R_1^s,T_2^s$, and the field inhomogeneities $B_0$ and $B_1^+$ (9 total parameters).
This hybrid-state sequence is optimized to minimize the CRB for $m_0^s,R_1^f,R_2^f,R_\text{x},R_1^s$, and $T_2^s$ in individual 4s long cycles with antiperiodic boundary conditions\cite{Asslander-hsCommPhysics,Asslander2023}. We use 3D radial koosh-ball k-space sampling with a 2D golden means pattern\cite{Chan2009} reshuffled to minimize eddy current artifacts\cite{Flassbeck2021}.

For our second application, we consider the MRF-FISP sequence\cite{Jiang2015a} designed to estimate $M_0$ and a single compartment $T_1,T_2$, which we implemented with 2D golden-angle radial sampling, 1ms BWTP=4 sinc-pulses, TR=10ms, TE=5ms, TI=20ms, and 10s for recovery to thermal equilibrium between 3 RF pattern repetitions (20.4s per IR curve).

\subsection{Data Simulation}
\label{subsec:datasim}

\begin{table}[tbp]
\begin{threeparttable}
    \caption{Parameter distributions used for dictionary simulation \label{tab:simvals}}
    \centering
    \begin{tabular}{ |c|c|c|c| }
        \hline
        \rowcolor[gray]{.7} qMT$^\text{a}$ & GM+WM (80\%) & Fat (10\%) & CSF (10\%)\\
        \hline
        $m_0^s$ & $\mathcal{N}(0.2, 0.2)_0$ & $\mathcal{N}(0.1, 0.1)_0$ & 0 \\ \hline
        $R_1^f$ (1/s) & $1/\mathcal{N}(3, 2)$ & $1/\mathcal{N}(0.4, 0.075)$ & $1/\mathcal{N}(4, 0.5)$ \\ \hline
        $R_2^f$ (1/s) & $\mathcal{N}(15, 10)$ & $1/\mathcal{N}(0.1, 0.020)$ & $1/\mathcal{N}(2, 0.25)$ \\ \hline
        $R_\text{x}  $ (1/s) & $\mathcal{N}(30, 10)$ & $\mathcal{N}(30, 10)$ & - \\ \hline
        $R_1^s$ (1/s) & $\mathcal{N}(4, 2)$ & $\mathcal{N}(4, 2)$ & - \\ \hline
        $T_2^s$ ($\upmu$s) & $\mathcal{N}(10, 3)_5^{20}$ & $\mathcal{N}(10, 3)_5^{20}$ & - \\ \hline
        $B_0$ & \multicolumn{3}{c|}{$\mathcal{U}_{[-\pi/\mathrm{TR},\pi/\mathrm{TR}]}$} \\ \hline
        $B_1^+$ & \multicolumn{3}{c|}{$\mathcal{N}(0.9, 0.3)_{0.6}^{1.2}$}\\\hline
        \multicolumn{4}{c}{~} \\ \hline
        \rowcolor[gray]{0.7} MRF-FISP$^\text{b}$ & GM+WM & Fat & CSF \\ \hline
        $T_1$ (ms) & 500:2:1500 & 250:2.4:550 & 3000:16:5000 \\ \hline
        $T_2$ (ms) & 10:0.38:200 & 60:0.65:140 & 1500:8.1:2500 \\ \hline
        $B_1^+$ & \multicolumn{3}{c|}{$1$} \\ \hline
    \end{tabular}
    \begin{tablenotes}
        \item[a] We use three Gaussian distributions, denoted by $\mathcal{N}(m,s)$ with mean $m$ and standard deviation $s$, corresponding to typical values in grey and white matter, fat, and cerebrospinal fluid at 3T\cite{Asslander2023,Bojorquez2017,Stanisz2005,Jiang2015a}. 
        The scripts denote truncation limits and $\mathcal{U}$ denotes a uniform distribution. The percentile brackets denote the relative size of each tissue type. 
        \item[b] MRF-FISP values are simulated on a discrete grid (min:stepsize:max).
    \end{tablenotes}
\end{threeparttable}
\end{table}

For the qMT model, we simulated a dictionary of approximately 600,000 fingerprints with the generalized Bloch framework\cite{Asslander2021} using three Gaussian distributions representing brain tissue (WM+GM), fat, and cerebrospinal fluid (CSF), as outlined in Table~\ref{tab:simvals}. 
67\% and 33\% of the total fingerprints were used for basis calculation and testing, respectively.

For MRF-FISP, we used Bloch simulations to compute a dictionary of 281,250 fingerprints over the Cartesian grid shown in Tab.~\ref{tab:simvals}, 
accounting for spoiling across the slice profile \cite{Ma2017} by taking the complex average of 1324 isochromats (5300 for cerebrospinal fluid)\cite{Malik2016}.

\subsection{Basis Optimization}
\label{subsec:basisopt}
For the qMT model, we consider 6 parameters of interest ($N_p=6$): $m_0^s,R_1^f,R_2^f,R_\text{x},R_1^s$, and $T_2^s$, and treat $M_0$, $B_0$, and $B_1^+$ as nuisance parameters which are only used indirectly in constructing $\mathbf{J}_\perp$ to calculate the orthogonalized derivatives for the parameters of interest.
For MRF-FISP, we consider $T_1$ and $T_2$ ($N_p=2$) while treating $M_0$ as a nuisance parameter. The signal derivatives for a given fingerprint are orthogonalized with respect to one another using QR factorization.

For the hybrid-state pulse sequence analyzed here, a sign change of $B_0$ entails simply a complex conjugation of the signal. Consequently, we expect real-valued bases with a symmetric $B_0$ distribution. To avoid approximation errors when randomly drawing samples, we complement each fingerprint with its complex conjugate to ensure that this symmetry is not broken and that we obtain real-valued bases.

For both applications, we perform the CRB-SVD using 10 evenly spaced values $\lambda\in[0,1)$, omitting $\lambda=1$ due to poor signal fidelity leading to severe image artifacts. SVDs were performed with 20GB memory distributed over 10 CPU threads (Intel Skylake 6148, Santa Clara, California, USA) for MRF-FISP and 700GB memory/30 CPU threads for the qMT model on our institution's computational cluster.
The latter took an average of 7 hours for each value of $\lambda$.

To evaluate the CRB-SVD method, we compare $B_{ac}$ (Eq.~\eqref{eq:acrb}) to $B_{ec}$ (Eq.~\eqref{eq:ccrb}) as a function of $N_c$ and $\lambda$. We also define the ratio
\begin{equation}
    R \triangleq \frac{1}{N_sN_p} \sum_{i=1}^{N_sN_p} \frac{B_{ac}(\boldsymbol\theta_i)}{B_{ec}(\boldsymbol\theta_i)}
    \label{eq:ratio}
\end{equation}
as a proxy for the optimality of the CRB-SVD basis. While the CRB-SVD computes a globally optimal solution for preserving $B_{ac}$, this solution is close to globally optimal for preserving $B_{ec}$ only when $R\approx1$. Let us denote the CRB-SVD solution with the values $B_{ac}^0$ and $B_{ec}^0$, and the values for a global minimizer of $B_{ec}$ with $B_{ec}^1$ and $B_{ac}^1$. Then, since $B_{ac}^0 \leq B_{ac}^1 \leq B_{ec}^1 \leq B_{ec}^0$, $R=B_{ac}^0/B_{ec}^0$ provides a tight bound on $B_{ec}^1$, the exact compressed CRB we would ideally like to minimize. Note $R\leq1$ by Eq.~\eqref{eq:inequality}.

\subsection{Simulation Experiments and Parameter Fitting}
\label{subsec:simexp}
To evaluate the impact of the CRB-SVD basis on parameter fitting, for both our test applications described in Section~\ref{subsec:sequences} we evaluate the bias and standard deviation of white matter parameter estimates across 1000 noisy measurements. For the qMT model we used the values $m_0^s=0.2,R_1^f=0.52/s,R_2^f=12.9/s,R_\text{x}=16.5/s,R_1^s=2.97/s$, and $T_2^s=12.4\mu s$\cite{Asslander2023}, while for MRF-FISP we used $T_1=810$ms and $T_2=25$ms. The signal was simulated using the same methods described in Section \ref{subsec:datasim} and compressed using the traditional SVD and CRB-SVD bases. We added Gaussian noise with an assumed $\text{SNR} \triangleq |M_0|/\sigma = 50$. 

For the qMT model, we used the non-linear least squares (NLLS) estimator using the Levenberg-Marquardt algorithm \cite{Marquardt1963} initialized with the ground-truth values.
For MRF-FISP, we used dictionary matching over the same Cartesian grid described in Section \ref{subsec:datasim}, where the stepsize was chosen to be smaller than the expected noise in the parameters to better approximate maximum likelihood estimation. Dictionary elements were compressed to the same subspace and matching was performed directly on the compressed coefficients\cite{McGivney2014}.

\subsection{In Vivo Imaging Experiments}
For the qMT sequence, we scanned the whole brain of a healthy subject on a 3T Biograph mMR (Siemens, Erlangen, Germany) with a 32-channel head coil and a 256mm isotropic FOV. To facilitate reconstruction with large $N_c$ values---which are used for validation and have comparably high memory demands---we used a lower-than-usual spatial resolution of 1.6mm isotropic resolution and 30 cycles of the hybrid-state sequence (similar to a multishot acquisition) for 12.6 min scan time. For the MRF-FISP sequence, we acquired a single slice through another healthy volunteer's brain on a 3T Prisma MRI scanner (Siemens, Erlangen, Germany). We used a 32-channel head coil, an FOV=256mm x 256mm, voxel size 1x1x4mm, and 3 cycles (3 radial spokes per frame), for approximately 1 min of scan time. Informed consent was obtained for both subjects in agreement with our IRB's requirements.

For both sequences, we reconstructed coefficient images for various combinations of $\{\lambda,N_c\}$ in \textit{Julia}\cite{Knopp2021} following the low-rank inversion approach\cite{McGivney2014,Asslander2018}. We used a Toeplitz approximation of the non-uniform FFT\cite{Wajer2001,Baron2018} for computational efficiency with non-Cartesian data and sensitivity encoding using coil maps calculated with ESPIRiT\cite{Uecker2014}. For the qMT sequence, we used a fixed locally low-rank penalty \cite{Trzasko2011,Zhang2015} strength and 250 iterations of FISTA\cite{Beck2009} to suppress artifacts and noise, followed by voxel-wise NLLS fitting with a maximum of 500 iterations. For MRF-FISP we used the conjugate gradient algorithm followed by voxel-wise dictionary matching as described in Section~\ref{subsec:simexp}.


\section{Results}
\label{sec:results}

\begin{figure}[tbp]
    \includegraphics[width=0.95\columnwidth]{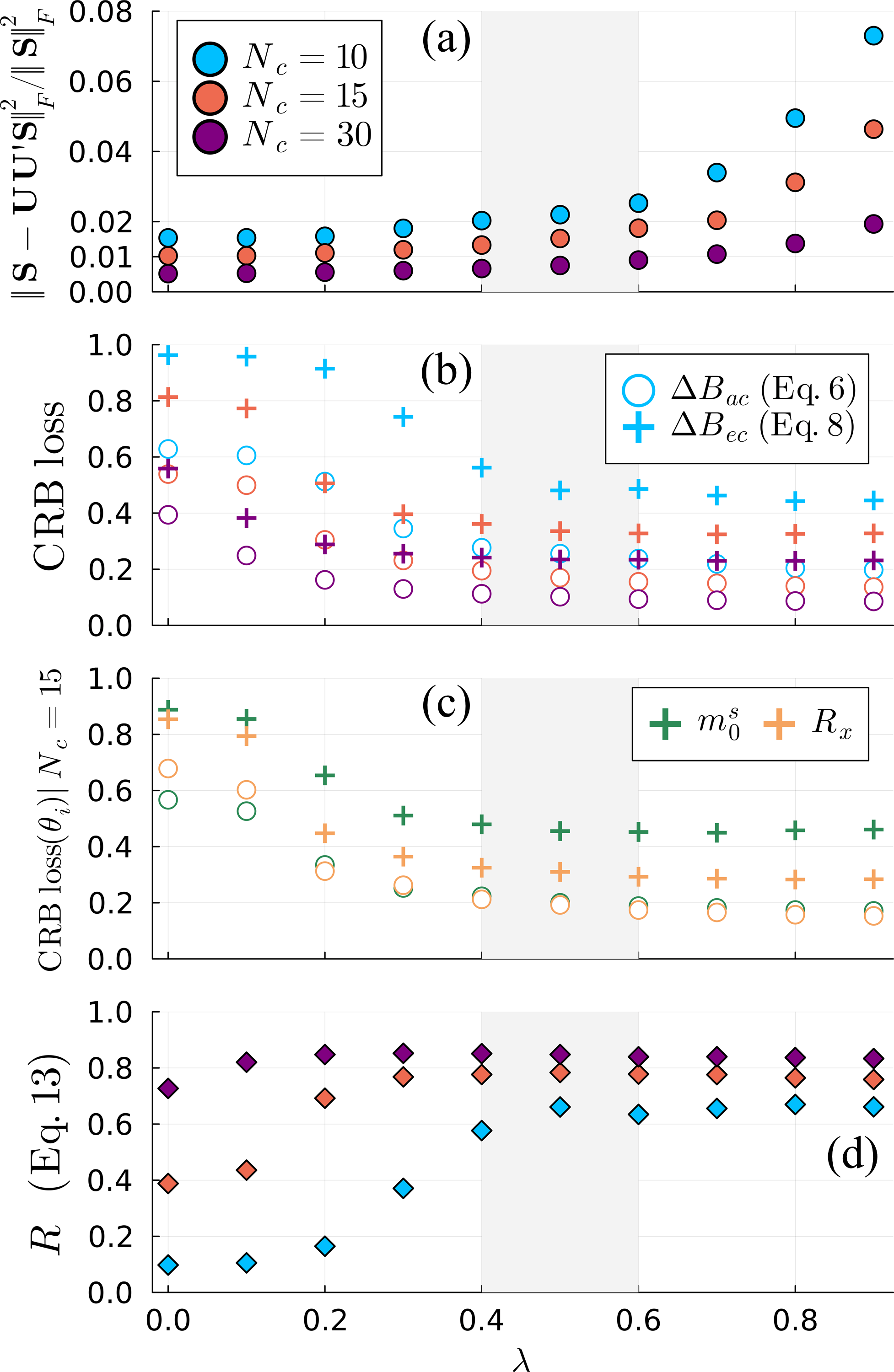}
    \caption{(a) Signal energy loss vs $\lambda$ normalized by the total signal energy in the qMT test dataset. (b) Cram\'er-Rao bound (CRB) loss calculated using the approximate compressed CRB ($\Delta B_{ac}$; Eq.~\eqref{eq:acrbloss}) and the exact compressed CRB ($\Delta B_{ec}$; Eq.~\eqref{eq:ccrbloss}), averaged over all fingerprints and orthogonalized derivatives in the test dataset. (c) CRB loss for $m_0^s$ and $R_\text{x}$ individually at $N_c=15$.
    (d) Average ratio of approximate compressed CRB to exact compressed CRB over the test dataset (Eq.~\eqref{eq:ratio}). The $\lambda$ values shaded in gray offer significant CRB improvements at a small cost to the signal fidelity for all $N_c$.}
    \label{fig:loss_split_mt}
\end{figure}


\begin{figure}[tbp]
    \centering
    \includegraphics[width=0.95\columnwidth]{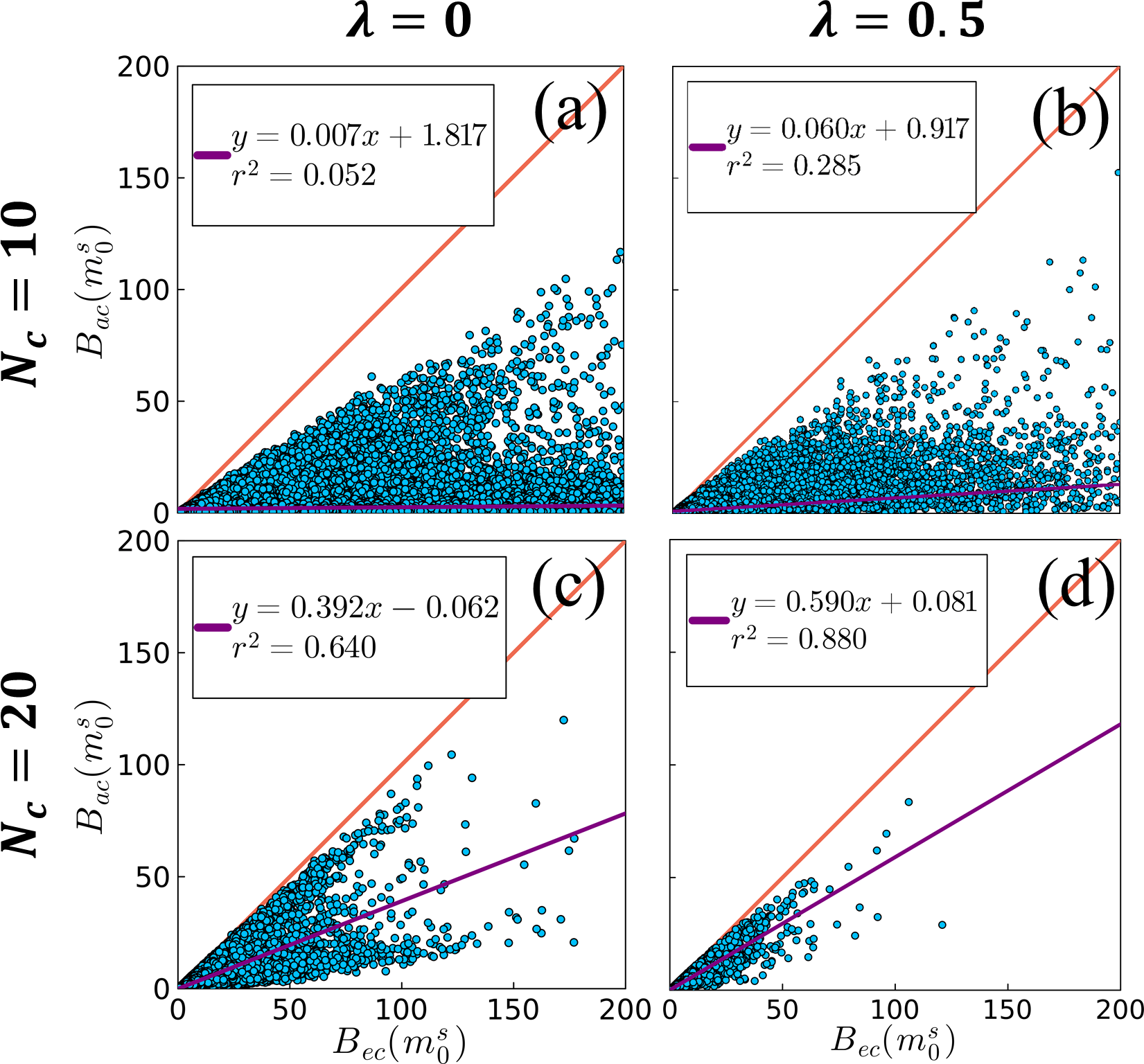}
    \caption{Representative scatter plots of the approximate compressed CRB of the semi-solid spin pool fraction ($B_{ac}(m_0^s)$) vs the exact compressed CRB ($B_{ec}(m_0^s)$) for different numbers of coefficients ($N_c$) and CRB-loss weightings ($\lambda$), where each dot represents a fingerprint in the qMT test set. Note each subplot's axes are limited to the same range for improved visualization. Linear regressions across all data points in each subplot (including those not plotted within the axis limits) are shown in purple in reference to the identity line in red. Correlations between $B_{ac}$ and $B_{ec}$ are improved with increasing $N_c$ and $\lambda$.}
    \label{fig:scatter_mt}
\end{figure}

\begin{figure}[tbp]
    \centering
    \includegraphics[width=\columnwidth]{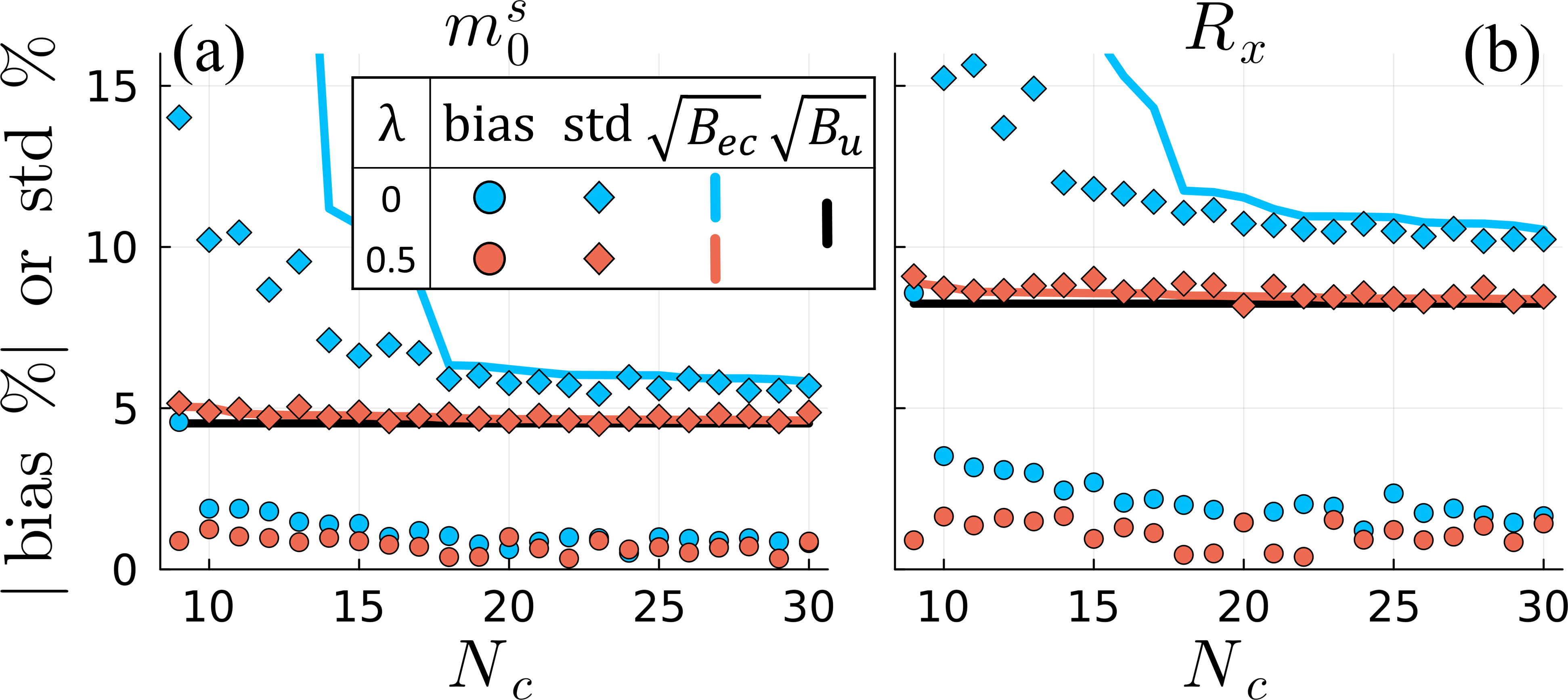}
    \caption{Bias (a--b) and standard deviation (c--d) of non-linear least squares (NLLS)-based $m_0^s$ and $R_\text{x}$ estimates of a typical white matter fingerprint derived from the traditional SVD ($\lambda=0$; blue) and CRB-SVD ($\lambda=0.5$; red) bases as a function of the number of coefficients $N_c$. Both metrics are normalized by the ground-truth $m_0^s$ and $R_\text{x}$ values, and the reference lines in (c--d) indicate the uncompressed CRB (Eq.~\eqref{eq:tcrb}) and exact compressed CRB (Eq.~\eqref{eq:ccrb}). Lower bias and variance more closely resembling minimum variance unbiased estimation is observed with the CRB-SVD basis, particularly for $N_c<18$, where the traditional SVD basis performs especially poorly.}
    \label{fig:biasvariance_m0sR1s_mt}
\end{figure}

\subsection{Quantitative Magnetization Transfer}
Fig.~\ref{fig:loss_split_mt} shows that increasing CRB-weightings ($\lambda$) leads to an improvement in CRB preservation at some cost to the preserved signal energy.
This improvement is more pronounced for small $N_c$: from $\lambda=0$ to $\lambda=0.5$ for $N_c=15$, the average $\Delta B_{ec}$ decreases from 0.81 to 0.34---equating to a 72\% decrease in the average CRB across all parameters---which notably is better than the preserved CRB for $\{\lambda=0,N_c=30\}$.
$\lambda \in \left[0.4,0.6\right]$ provides nearly the maximal improvement in the CRB with minimal cost to the preserved energy for all $N_c$. We observe that the $R$ values (Eq.~\eqref{eq:ratio}) achieved by the CRB-SVD bases have a cap of 0.85 for $N_c=30$---the limit of a computationally achievable reconstruction. This indicates the practical infeasibility of capturing the entire span of the signal derivatives within a compact basis for this application but also suggests good practical utility for the CRB-SVD basis: the traditional SVD basis for $N_c=15$ has $R=0.39$, demonstrating its suboptimality in preserving $B_{ec}$.

\begin{figure*}[tbp]
    \centering
    \includegraphics[width=\textwidth]{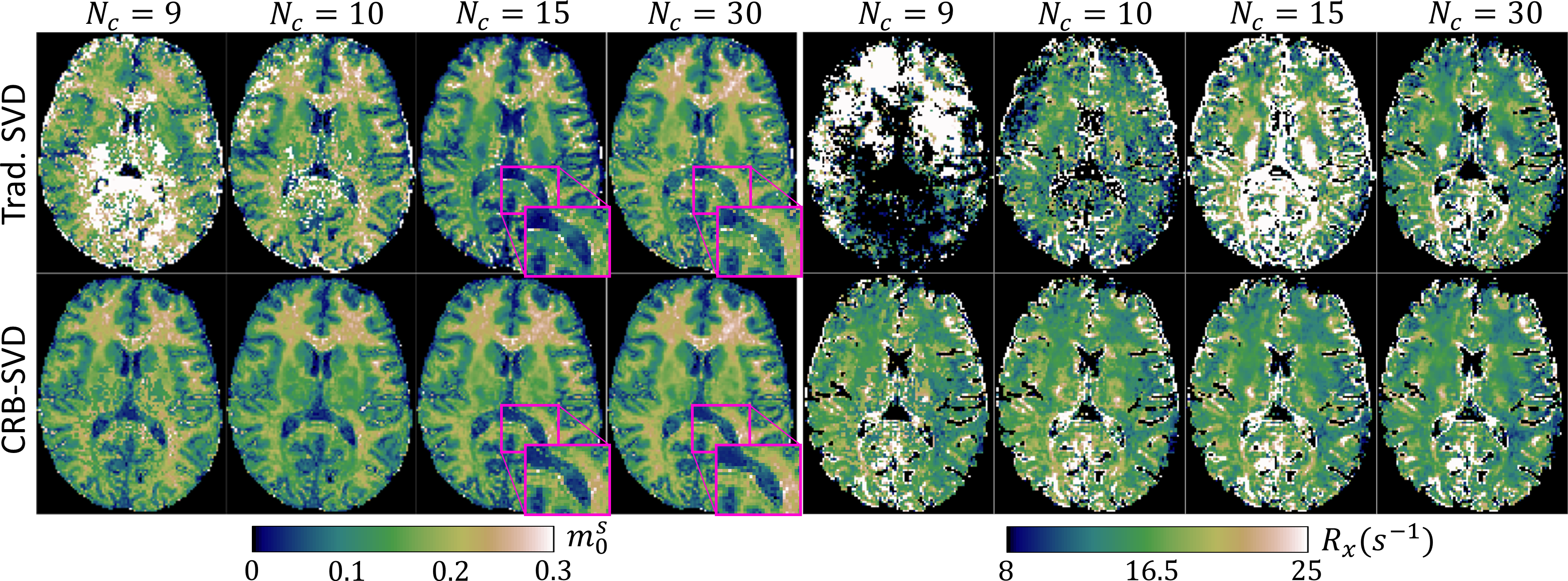}
    \caption{Comparison of non-linear least squares fits of the semi-solid spin pool's fractional size $m_0^s$ (a) and magnetization exchange rate $R_\text{x}$ (b) extracted from the traditional SVD ($\lambda=0$) and proposed CRB-SVD basis ($\lambda=0.5$) for different numbers of coefficients ($N_c$), where $N_c=30$ serves as a gold standard. The CRB-SVD basis improves the precision of the $m_0^s$ (magnifications) and the accuracy of the $R_\text{x}$ maps for small $N_c$.
    }
    \label{fig:maps_mt}
\end{figure*}

While $B_{ac}$ never quite approaches $B_{ec}$, Fig.~\ref{fig:scatter_mt} investigates this approximation by plotting a regression across all test fingerprints. Note that all points lie below the reference identity line, consistent with Eq.~\eqref{eq:inequality}. $B_{ac}$ correlates poorly with $B_{ec}$ for $N_c=10$---which is close to the number of parameters in this model---where
deviations from the identity line are generally more severe for large CRB values.
The correlation improves with increasing $\lambda$, and even more so with increasing $N_c$---which also significantly reduces the CRB, such that all test points are captured within the plotted axis limits.

Fig.~\ref{fig:biasvariance_m0sR1s_mt} shows that the bias and variance of NLLS-based $m_0^s$ and $R_\text{x}$ estimates are reduced with the CRB-SVD ($\lambda=0.5$) basis compared to the traditional SVD ($\lambda=0$) for all $N_c$.
The difference is particularly pronounced for $N_c<18$, where the $B_{ec}$ reference shows that the estimation problem becomes particularly difficult with the traditional SVD basis (blue line) and NLLS deviates significantly from a minimum variance unbiased estimator. In particular, the variance is reduced relative to $B_{ec}$ at the cost of introducing significant bias. As there is no model mismatch in this simulation, this bias reflects a failure of the estimator rather than the biophysical model. However, both bias and variance remain stable for the CRB-SVD basis across all $N_c$ and both parameters.

In vivo, we similarly observe improved performance consistency using the CRB-SVD basis across all $N_c$ as shown in Fig.~\ref{fig:maps_mt}, where $N_c=30$ is used as a gold standard.
The magnifications demonstrate the improved precision in $m_0^s$, most prominently for small $N_c$.
In contrast to the improved precision in $m_0^s$, the maps of the generally worse-conditioned $R_\text{x}$ parameter reveal mostly improvements in accuracy. The increasing CRB of $R_\text{x}$ for $N_c\leq18$ results in a significant bias with the traditional SVD basis which is not seen with the use of the CRB-SVD basis. Here, increasing $N_c$ is of limited benefit, and increasing $\lambda$ has a much greater impact on accuracy. Since $R_\text{x}$ is a comparably difficult-to-estimate parameter, the accuracy and precision of its estimate tend to guide the selection of the appropriate $N_c$ for this application.


\subsection{MRF-FISP}
\begin{figure}[tbp]
    \centering
    \includegraphics[width=\columnwidth]{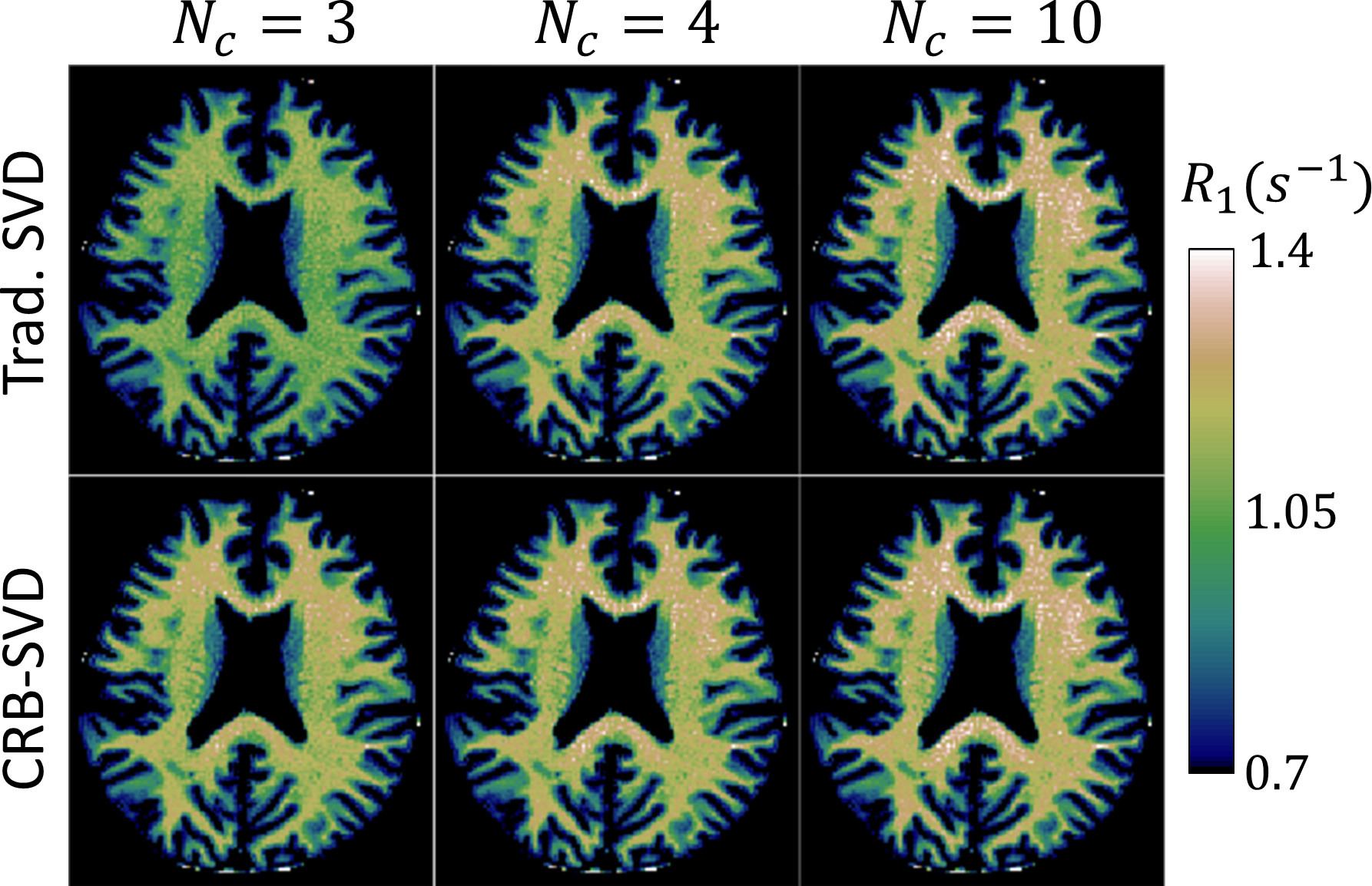}
    \caption{Comparison of $R_1=1/T_1$ maps extracted from the MRF-FISP data that were reconstructed with a traditional SVD basis ($\lambda=0$) and the proposed CRB-SVD basis ($\lambda=0.3$) for different subspace sizes ($N_c$), where $N_c=10$ serves as a gold standard.
    The CRB-SVD basis offers improved parameter accuracy at $N_c=3$, the number of parameters in the biophysical model, and is non-inferior for $N_c\geq4$.}
    \label{fig:dm_fisp}
\end{figure}

Fig.~\ref{fig:dm_fisp} compares in vivo fits of $R_1$ derived from the traditional SVD and CRB-SVD bases with dictionary matching. Here, we observe that the CRB-SVD yields similar results for $N_c\geq4$, with a benefit seen mostly at the extreme of $N_c=3$ where the traditional SVD's estimates are substantially biased.

\section{Discussion} \label{sec:disc}
Our results suggest that performing the SVD on the signals alone is suboptimal in preserving the CRB of the model's parameters, which depends on preserving the orthogonal components of the signal's derivatives. In this work, we proposed a method for incorporating CRB preservation into an SVD-based calculation of the basis. This approach utilizes a geometric interpretation of the CRB \cite{Scharf1993} and relies on an approximation of the compressed CRB. In general, we find that a stronger CRB weighting (increased $\lambda$) results in improved CRB preservation with a small cost to signal fidelity. Additionally, we find that a stronger CRB weighting also promotes a better approximation of the exact compressed CRB ($B_{ec}$) by the approximate compressed CRB ($B_{ac}$), improving confidence in the quality of the solution. In simulation, we found that when the number of measurements ($N_c$) was similar to the number of model parameters, the proposed method reduced both bias and variance of the parameter estimates, which will be further discussed in Section~\ref{subsec:disc_acc}. In vivo, the CRB-SVD basis showed some improvements at minimal $N_c$ when analyzing the MRF-FISP sequence and non-inferiority otherwise. For the qMT sequence, the proposed approach offered improved accuracy and precision at constant $N_c$, or the ability to reduce $N_c$ while retaining accuracy and precision.

While we focused here on adding a CRB preservation term to the traditional SVD basis optimization objective, we emphasize that signal fidelity is still of fundamental importance.
In the image reconstruction process, unrepresented signal energy causes inconsistency with the measured data, leading to artifacts scattered throughout the coefficient images that can bias the parameter estimates. Still, we have shown in this article that there is a range of $\lambda$-values where the improved CRB outweighs the small cost in signal fidelity. This suggests that the slight increase in artifacts is less important than the improvements to the parameter estimation optimization landscape gained from reduced CRBs.

\subsection{Practical Considerations}
The proposed CRB-SVD method involves a simple modification to the traditional SVD basis calculation, requiring only precomputation and orthogonalization of the signal's derivatives with respect to the model parameters. These can be calculated analytically for many qMRI sequences, but automatic differentiation\cite{Lee2019,Rakshit2021} or physics-inspired neural networks\cite{Cuomo2022} can also be used.
The appropriate $\lambda$ is application dependent, where a slightly higher $\lambda$ was found to be suitable for the qMT application because the improved CRB, especially at low subspace sizes, outweighed the loss in preserved signal. We postulate that the selection of $\lambda$ may, in general, also be sensitive to the signal variety---e.g. the variety of anatomy and tissue types---that needs to be accurately represented in the basis.
Regardless, a moderate $\lambda$ value of 0.3--0.5 generally appears to be a good starting point for other applications.
Once computed, the CRB-SVD bases can be directly utilized with standard reconstruction pipelines, such as in BART\cite{BART} for typical subspace reconstruction tasks or even multitasking applications\cite{Christodoulou2018,Wang2022b,Ma2022} where the basis for some dimension (e.g., $T_1$ relaxation) can be predetermined.


The practical benefit for subspace reconstructions in terms of computational and memory demands is significant, which can be limiting particularly for non-Cartesian 4D+ applications. Many iterative reconstruction algorithms require the computation and storage of the normal operators, whose size scales quadratically with $N_c$. For example, in the qMT case, reducing 30 coefficients to 15 or even 10 reduces the memory requirements by a factor of 4 and 9, respectively. In practice, for a 160x150x120 matrix size (which corresponds to the full-brain coverage with 1.6mm isotropic resolution we used in this article)---assuming complex-valued single-precision floating points and 2x oversampling in each dimension for a Toeplitz approximation---is the difference between 39GB/17GB and 154GB memory, which is often unavailable on a typical workstation.
For 1.0mm isotropic resolution\cite{Asslander2023}, the reduction is from 675GB to 169GB or 75GB memory, conferring feasibility on most high-performance computational platforms.
A smaller number of coefficients also directly speeds up the calculation of the Toeplitz kernels and shortens the per-iteration computation time (the normal operators involve $\mathcal{O}(N_c^2)$ complex-valued floating point multiplications per k-space data point), reducing the overall reconstruction time (assuming a fixed number of iterations).

\subsection{Limitations and Drawbacks}
\label{subsec:limitations}
While general superiority or even non-inferiority of the CRB-SVD basis is difficult to prove and is dependent on the figure of merit, we analyzed two extreme cases here to provide information across a wide spectrum of potential applications. We analyzed the MRF-FISP sequence as it is known for its very compact representation\cite{McGivney2014}. For this reason, we did not expect major improvements and Fig.~\ref{fig:dm_fisp} largely confirms this expectation. At the other end of the spectrum, we analyzed a qMT sequence that has a less compact representation and showed that the CRB-SVD has substantial benefits in such a case. Many practical applications---and consequently the expected benefit---likely lie between these two illustrative examples. Nonetheless, we have identified other extreme cases from the literature: McGivney et. al. \cite{McGivney2014} used $N_c = 200$ to represent the signal of the original MRF approach\cite{Ma2013}, and for such cases, we expect a substantial benefit when using the CRB-SVD over its traditional counterpart. 

While the performance and optimality of the proposed CRB-SVD are straightforward to verify (c.f. Fig.~\ref{fig:loss_split_mt}), an important limitation is that we did not perform direct minimization of the exact compressed CRB. This would improve the optimality of the CRB-SVD bases with potential further improvements in representational efficiency (i.e., improved accuracy and precision at smaller subspace sizes). Direct minimization of the exact compressed CRB is not straightforward as the orthogonal projection onto $\langle \mathbf{J}_i \rangle$ depends on $\mathbf{U}$ (Eq.~\eqref{eq:ccrb}), and we need to additionally fulfill the orthonormal basis constraint ($\mathbf{U}'\mathbf{U}=\mathbf{I}$) because our CRB expression assumes that the basis coefficients have no cross-talk, and bases that are not normalized can artificially decrease the CRB. This is a non-convex minimization over the complex Grassmannian likely requiring an iterative procedure, e.g. stochastic gradient descent, which will be the subject of future work.

In this article, we assumed a white Gaussian noise model and did not consider deviations under experimental conditions, e.g. due to parallel imaging\cite{Bouhrara2018} or compressed sensing. We note that the CRB is somewhat limited as a design criterion because it considers only local properties of the log-likelihood function and an optimistic CRB can sometimes mask poor model identifiability\cite{Nehorai2014}; i.e., that the data can be explained by multiple combinations of parameters\cite{Jelescu2016}. For example, a model with poor identifiability would have many local minima which might cause convergence issues in gradient-based estimators\cite{Fletcher2000}---which would have to be mitigated by the selection of appropriate constraints in parameter space. A Barankin-type bound, while more difficult to compute, would better capture the global structure of the statistical model and more accurately reflect realizable system performance in low SNR regimes\cite{McAulay1971} relevant in many qMRI applications, and is another avenue for future work.

One drawback of the proposed method is the memory requirements for performing the CRB-SVD, which as described in Section \ref{subsec:crbbases} scale linearly with the number of included orthogonalized derivatives and fingerprints. The sample size for the 9-parameter qMT model was dictated by the practical limitation of the maximum memory available on our computational cluster, though there was virtually no difference in the traditional SVD-derived basis with 5x the sample size.
Additionally, for most applications, the basis functions have to be calculated only once, while the memory requirements for the reconstruction have to be fulfilled for every dataset. In future work, we will investigate more memory-efficient algorithms for performing large-scale SVDs\cite{Yang2018}.

\subsection{Cram\'er-Rao Bound: Precision or Accuracy?}
\label{subsec:disc_acc}
The CRB is most often associated with the variance or noise of a parameter estimate.
However, maximum likelihood estimators (e.g., the commonly used dictionary matching) are only asymptotically unbiased (or $\sqrt{N_c}$-consistent), i.e., for many measurements\cite{Kay1993,Newey1994}. Least squares estimators (e.g., NLLS) are asymptotically unbiased only for white Gaussian noise models\cite{Kay1993,Wu1981}. In vivo, where acquisition time is a major constraint, only limited measurements can be made. In qMRI, the number of measurements has often been similar to the number of model parameters ($N_c\approx N_p$ in the context of subspace reconstruction) tracing back to the original DESPOT method\cite{Deoni2003}, which results in non-negligible bias. Unmodeled biophysical effects (e.g. unmodeled tissue compartments or diffusion in our qMT model\cite{Asslander2022}), regularization in the image reconstruction\cite{Fessler2002}, and image artifacts all inexorably introduce further errors into the parameter estimation process. As a consequence, we practically cannot expect unbiased parameter estimation and the performance predicted by the CRB is rarely obtained. Nonetheless, the CRB is commonly used as a proxy for the ``SNR-efficiency'' or ``conditioning'' of the qMRI system\cite{Zhao2019,Haldar2019}, and our work adopts this heuristic. Our simulations and in vivo experiments confirm that dictionary matching and NLLS estimates of the parameters are biased when $N_c\approx N_p$, and particularly so with the traditional SVD basis, limiting the linear compressibility of advanced quantitative models that can practically be achieved. However, our results show that the use of the CRB-SVD basis improves both precision and accuracy in subspace reconstruction.

The observations about precision and accuracy with respect to the theoretical CRB made in this article are not new in the qMRI literature, and many papers report significant differences in both the mean and standard deviation of estimated parameters for CRB-optimized sequences\cite{Teixeira2018,Zhao2019,Asslander-hsMRM,Liu2017,Lee2019,Nataraj2017a}. Notably, Zhao et. al. observe in simulated fully-sampled data an unbiased estimate of the relaxation times with the expected reduction in noise for a CRB-optimized sequence (note subspace reconstruction was not used in their work)\cite{Zhao2019}. However, they find in the (spirally) undersampled regime that the parameter estimates are biased and the CRB-optimized sequence helps to improve both the accuracy and precision of the parameter maps. Unfortunately, limited conclusions about the relationship between the CRB and observed bias can be drawn from other studies involving CRB-based optimization of the scan parameters that show in vivo results\cite{Teixeira2018,Asslander-hsMRM,Liu2017,Lee2019,Nataraj2017a} due to bias introduced by unmodeled biophysical effects such as magnetization transfer.

\section{Conclusion}
\label{sec:conc}
We present a method to incorporate CRB preservation in addition to signal fidelity in the subspace basis optimization objective. We approximate the compressed domain CRB with a computationally efficient alternative, which yields a sufficiently optimal solution in practice. The proposed CRB-SVD basis is a drop-in replacement for the traditional SVD basis with improved representational compactness, promoting improved parameter accuracy and precision at smaller subspace sizes and offering computational speedups and memory savings for subspace reconstruction tasks.

\bibliographystyle{IEEEtran}
\bibliography{references}

\begin{thebibliography}{10}
\providecommand{\url}[1]{#1}
\csname url@samestyle\endcsname
\providecommand{\newblock}{\relax}
\providecommand{\bibinfo}[2]{#2}
\providecommand{\BIBentrySTDinterwordspacing}{\spaceskip=0pt\relax}
\providecommand{\BIBentryALTinterwordstretchfactor}{4}
\providecommand{\BIBentryALTinterwordspacing}{\spaceskip=\fontdimen2\font plus
\BIBentryALTinterwordstretchfactor\fontdimen3\font minus \fontdimen4\font\relax}
\providecommand{\BIBforeignlanguage}[2]{{%
\expandafter\ifx\csname l@#1\endcsname\relax
\typeout{** WARNING: IEEEtran.bst: No hyphenation pattern has been}%
\typeout{** loaded for the language `#1'. Using the pattern for}%
\typeout{** the default language instead.}%
\else
\language=\csname l@#1\endcsname
\fi
#2}}
\providecommand{\BIBdecl}{\relax}
\BIBdecl

\bibitem{Deoni2003}
S.~C. Deoni, B.~K. Rutt, and T.~M. Peters, ``{Rapid combined T1 and T2 mapping using gradient recalled acquisition in the steady state},'' \emph{Magnetic Resonance in Medicine}, vol.~49, no.~3, pp. 515--526, 2003.

\bibitem{Hennig1988}
J.~Hennig, ``{Multiecho imaging sequences with low refocusing flip angles},'' \emph{Journal of Magnetic Resonance (1969)}, vol.~78, no.~3, pp. 397--407, 1988.

\bibitem{Look1970}
D.~C. Look and D.~R. Locker, ``{Time saving in measurement of NMR and EPR relaxation times},'' \emph{Review of Scientific Instruments}, vol.~41, no.~2, pp. 250--251, 1970.

\bibitem{Ma2013}
D.~Ma, V.~Gulani, N.~Seiberlich, K.~Liu, J.~L. Sunshine, J.~L. Duerk, and M.~A. Griswold, ``{Magnetic resonance fingerprinting},'' \emph{Nature}, vol. 495, no. 7440, pp. 187--192, 2013.

\bibitem{Christodoulou2018}
A.~G. Christodoulou, J.~L. Shaw, C.~Nguyen, Q.~Yang, Y.~Xie, N.~Wang, and D.~Li, ``{Magnetic resonance multitasking for motion-resolved quantitative cardiovascular imaging},'' \emph{Nature Biomedical Engineering}, vol.~2, no.~4, pp. 215--226, 2018.

\bibitem{Wang2019epti}
F.~Wang, Z.~Dong, T.~G. Reese, B.~Bilgic, M.~{Katherine Manhard}, J.~Chen, J.~R. Polimeni, L.~L. Wald, and K.~Setsompop, ``{Echo planar time-resolved imaging (EPTI)},'' \emph{Magnetic Resonance in Medicine}, vol.~81, no.~6, pp. 3599--3615, 2019.

\bibitem{Wang2022}
F.~Wang, Z.~Dong, T.~G. Reese, B.~Rosen, L.~L. Wald, and K.~Setsompop, ``{3D Echo Planar Time-resolved Imaging (3D-EPTI) for ultrafast multi-parametric quantitative MRI},'' \emph{NeuroImage}, vol. 250, no. December 2021, p. 118963, 2022.

\bibitem{Sbrizzi2018}
A.~Sbrizzi, O.~van~der Heide, M.~Cloos, A.~van~der Toorn, H.~Hoogduin, P.~R. Luijten, and C.~A. van~den Berg, ``{Fast quantitative MRI as a nonlinear tomography problem},'' \emph{Magnetic Resonance Imaging}, vol.~46, no. June 2017, pp. 56--63, 2018.

\bibitem{Asslander-hsCommPhysics}
J.~Assl{\"{a}}nder, D.~S. Novikov, R.~Lattanzi, D.~K. Sodickson, and M.~A. Cloos, ``{Hybrid-state free precession in nuclear magnetic resonance},'' \emph{Communications Physics}, vol.~2, no.~1, 2019.

\bibitem{Liang2007}
Z.-P. Liang, ``{Spatiotemporal Imaging with Partially Separable Functions},'' \emph{2007 Joint Meeting of the 6th International Symposium on Noninvasive Functional Source Imaging of the Brain and Heart and the International Conference on Functional Biomedical Imaging}, vol.~2, pp. 181--182, oct 2007.

\bibitem{Huang2012}
C.~Huang, C.~G. Graff, E.~W. Clarkson, A.~Bilgin, and M.~I. Altbach, ``{T 2 mapping from highly undersampled data by reconstruction of principal component coefficient maps using compressed sensing},'' \emph{Magnetic Resonance in Medicine}, vol.~67, no.~5, pp. 1355--1366, may 2012.

\bibitem{McGivney2014}
D.~F. McGivney, E.~Pierre, D.~Ma, Y.~Jiang, H.~Saybasili, V.~Gulani, and M.~A. Griswold, ``{SVD compression for magnetic resonance fingerprinting in the time domain},'' \emph{IEEE Transactions on Medical Imaging}, vol.~33, no.~12, pp. 2311--2322, 2014.

\bibitem{Tamir2017}
J.~I. Tamir, M.~Uecker, W.~Chen, P.~Lai, M.~T. Alley, S.~S. Vasanawala, and M.~Lustig, ``{T 2 shuffling: Sharp, multicontrast, volumetric fast spin‐echo imaging},'' \emph{Magnetic Resonance in Medicine}, vol.~77, no.~1, pp. 180--195, jan 2017.

\bibitem{Asslander2018}
J.~Assl{\"{a}}nder, M.~A. Cloos, F.~Knoll, D.~K. Sodickson, J.~Hennig, and R.~Lattanzi, ``{Low rank alternating direction method of multipliers reconstruction for MR fingerprinting},'' \emph{Magnetic Resonance in Medicine}, vol.~79, no.~1, pp. 83--96, jan 2018.

\bibitem{Zhao2015}
B.~Zhao, W.~Lu, T.~K. Hitchens, F.~Lam, C.~Ho, and Z.~P. Liang, ``{Accelerated MR parameter mapping with low-rank and sparsity constraints},'' \emph{Magnetic Resonance in Medicine}, vol.~74, no.~2, pp. 489--498, 2015.

\bibitem{Zhao2018}
B.~Zhao, K.~Setsompop, E.~Adalsteinsson, B.~Gagoski, H.~Ye, D.~Ma, Y.~Jiang, P.~{Ellen Grant}, M.~A. Griswold, and L.~L. Wald, ``{Improved magnetic resonance fingerprinting reconstruction with low-rank and subspace modeling},'' \emph{Magnetic Resonance in Medicine}, vol.~79, no.~2, pp. 933--942, feb 2018.

\bibitem{Hennig1991}
J.~Hennig, ``{Echoes—how to generate, recognize, use or avoid them in MR-imaging sequences. Part I: Fundamental and not so fundamental properties of spin echoes},'' \emph{Concepts in Magnetic Resonance}, vol.~3, no.~3, pp. 125--143, jul 1991.

\bibitem{Weigel2015}
M.~Weigel, ``{Extended phase graphs: Dephasing, RF pulses, and echoes - Pure and simple},'' \emph{Journal of Magnetic Resonance Imaging}, vol.~41, no.~2, pp. 266--295, 2015.

\bibitem{Malik2018}
S.~J. Malik, R.~P.~A. Teixeira, and J.~V. Hajnal, ``{Extended phase graph formalism for systems with magnetization transfer and exchange},'' \emph{Magnetic Resonance in Medicine}, vol.~80, no.~2, pp. 767--779, 2018.

\bibitem{Haldar2009}
J.~P. Haldar, D.~Hernando, and Z.~P. Liang, ``{Super-Resolution reconstruction of mr image sequences with contrast modeling},'' \emph{Proceedings - 2009 IEEE International Symposium on Biomedical Imaging: From Nano to Macro, ISBI 2009}, no.~1, pp. 266--269, 2009.

\bibitem{Zhao2016}
B.~Zhao, K.~Setsompop, H.~Ye, S.~F. Cauley, and L.~L. Wald, ``{Maximum Likelihood Reconstruction for Magnetic Resonance Fingerprinting},'' \emph{IEEE Transactions on Medical Imaging}, vol.~35, no.~8, pp. 1812--1823, 2016.

\bibitem{Wang2018mbir}
X.~Wang, V.~Roeloffs, J.~Klosowski, Z.~Tan, D.~Voit, M.~Uecker, and J.~Frahm, ``{Model-based T1 mapping with sparsity constraints using single-shot inversion-recovery radial FLASH},'' \emph{Magnetic Resonance in Medicine}, vol.~79, no.~2, pp. 730--740, 2018.

\bibitem{Scholand2023}
N.~Scholand, X.~Wang, V.~Roeloffs, S.~Rosenzweig, and M.~Uecker, ``{Quantitative MRI by nonlinear inversion of the Bloch equations},'' \emph{Magnetic Resonance in Medicine}, no. March, pp. 1--19, apr 2023.

\bibitem{Asslander2020}
J.~Assl{\"{a}}nder, ``{A Perspective on MR Fingerprinting},'' \emph{Journal of Magnetic Resonance Imaging}, pp. 1--10, 2020.

\bibitem{Shafieizargar2023}
B.~Shafieizargar, R.~Byanju, J.~Sijbers, S.~Klein, A.~J. den Dekker, and D.~H.~J. Poot, ``{Systematic review of reconstruction techniques for accelerated quantitative MRI.}'' \emph{Magnetic Resonance in Medicine}, no. April, pp. 1--37, 2023.

\bibitem{Jiang2015a}
Y.~Jiang, D.~Ma, N.~Seiberlich, V.~Gulani, and M.~A. Griswold, ``{MR fingerprinting using fast imaging with steady state precession (FISP) with spiral readout},'' \emph{Magnetic Resonance in Medicine}, vol.~74, no.~6, pp. 1621--1631, 2015.

\bibitem{Nataraj2018}
G.~Nataraj, J.~F. Nielsen, C.~Scott, and J.~A. Fessler, ``{Dictionary-Free MRI PERK: Parameter Estimation via Regression with Kernels},'' \emph{IEEE Transactions on Medical Imaging}, vol.~37, no.~9, pp. 2103--2114, 2018.

\bibitem{Virtue2018}
P.~Virtue, S.~X. Yu, and M.~Lustig, ``{Better than real: Complex-valued neural nets for MRI fingerprinting},'' \emph{Proceedings - International Conference on Image Processing, ICIP}, vol. 2017-Sept, pp. 3953--3957, 2018.

\bibitem{Cohen2018a}
O.~Cohen, B.~Zhu, and M.~S. Rosen, ``{MR fingerprinting Deep RecOnstruction NEtwork (DRONE)},'' \emph{Magnetic Resonance in Medicine}, vol.~80, no.~3, pp. 885--894, sep 2018.

\bibitem{Zhang2022}
X.~Zhang, Q.~Duchemin, K.~Liu*, C.~Gultekin, S.~Flassbeck, C.~Fernandez‐Granda, and J.~Assl{\"{a}}nder, ``{Cram{\'{e}}r–Rao bound‐informed training of neural networks for quantitative MRI},'' \emph{Magnetic Resonance in Medicine}, no. February, pp. 1--13, mar 2022.

\bibitem{Cramer}
H.~Cram\'{e}r, \emph{Mathematical Methods of Statistics}.\hskip 1em plus 0.5em minus 0.4em\relax Princeton University Press, 09 1946.

\bibitem{Jones1997}
J.~A. Jones, ``{Optimal Sampling Strategies for the Measurement of Relaxation Times in Proteins},'' \emph{Journal of Magnetic Resonance}, vol. 126, no.~2, pp. 283--286, 1997.

\bibitem{Teixeira2018}
R.~P.~A. Teixeira, S.~J. Malik, and J.~V. Hajnal, ``{Joint system relaxometry (JSR) and Cr{\'{a}}mer-Rao lower bound optimization of sequence parameters: A framework for enhanced precision of DESPOT T1 and T2 estimation},'' \emph{Magnetic Resonance in Medicine}, vol.~79, no.~1, pp. 234--245, 2018.

\bibitem{Zhao2019}
B.~Zhao, J.~P. Haldar, C.~Liao, D.~Ma, Y.~Jiang, M.~A. Griswold, K.~Setsompop, and L.~L. Wald, ``{Optimal experiment design for magnetic resonance fingerprinting: Cram{\'{e}}r-rao bound meets spin dynamics},'' \emph{IEEE Transactions on Medical Imaging}, vol.~38, no.~3, pp. 844--861, 2019.

\bibitem{Asslander-hsMRM}
J.~Assl{\"{a}}nder, R.~Lattanzi, D.~K. Sodickson, and M.~A. Cloos, ``{Optimized quantification of spin relaxation times in the hybrid state},'' \emph{Magnetic Resonance in Medicine}, vol.~82, no.~4, pp. 1385--1397, 2019.

\bibitem{Liu2017}
Y.~Liu, J.~R. Buck, and V.~N. Ikonomidou, ``{Generalized min-max bound-based MRI pulse sequence design framework for wide-range T1 relaxometry: A case study on the tissue specific imaging sequence},'' \emph{PLoS ONE}, vol.~12, no.~2, pp. 1--20, 2017.

\bibitem{Lee2019}
P.~K. Lee, L.~E. Watkins, T.~I. Anderson, G.~Buonincontri, and B.~A. Hargreaves, ``{Flexible and efficient optimization of quantitative sequences using automatic differentiation of Bloch simulations},'' \emph{Magnetic Resonance in Medicine}, vol.~82, no.~4, pp. 1438--1451, 2019.

\bibitem{Nataraj2017a}
G.~Nataraj, J.~F. Nielsen, and J.~A. Fessler, ``{Optimizing MR scan design for model-based T1,T2 estimation from steady-state sequences},'' \emph{IEEE Transactions on Medical Imaging}, vol.~36, no.~2, pp. 467--477, 2017.

\bibitem{Haldar2019}
J.~P. Haldar and D.~Kim, ``{OEDIPUS: An Experiment Design Framework for Sparsity-Constrained MRI},'' \emph{IEEE Transactions on Medical Imaging}, vol.~38, no.~7, pp. 1545--1558, 2019.

\bibitem{Whitaker2020}
S.~T. Whitaker, G.~Nataraj, J.~F. Nielsen, and J.~A. Fessler, ``{Myelin water fraction estimation using small-tip fast recovery MRI},'' \emph{Magnetic Resonance in Medicine}, vol.~84, no.~4, pp. 1977--1990, 2020.

\bibitem{Wang2023}
N.~Wang, X.~Cao, S.~S. Iyer, C.~Liao, P.~K. Lee, M.~Zhang, and K.~Setsompop, ``{Optimization of Magnetic Resonance Fingerprinting with Subspace Reconstruction},'' \emph{Proc. Intl. Soc. Mag. Reson. Med.}, 2023.

\bibitem{Jelescu2016}
I.~O. Jelescu, J.~Veraart, E.~Fieremans, and D.~S. Novikov, ``{Degeneracy in model parameter estimation for multi-compartmental diffusion in neuronal tissue},'' \emph{NMR in Biomedicine}, vol.~29, no.~1, pp. 33--47, 2016.

\bibitem{Hero1992}
A.~O. Hero, ``{A Cramer-Rao Type Lower Bound for Essentially Unbiased Parameter Estimation},'' US Dept of the Air Force, Tech. Rep., jan 1992.

\bibitem{Scharf1993}
L.~L. Scharf and L.~T. McWhorter, ``{Geometry of the Cramer-Rao bound},'' \emph{Signal Processing}, vol.~31, no.~3, pp. 301--311, apr 1993.

\bibitem{Henkelman1993}
R.~M. Henkelman, X.~Huang, Q.~Xiang, G.~J. Stanisz, S.~D. Swanson, and M.~J. Bronskill, ``{Quantitative interpretation of magnetization transfer},'' \emph{Magnetic Resonance in Medicine}, vol.~29, no.~6, pp. 759--766, 1993.

\bibitem{Helms2009}
G.~Helms and G.~E. Hagberg, ``{In vivo quantification of the bound pool T1 in human white matter using the binary spin-bath model of progressive magnetization transfer saturation},'' \emph{Physics in Medicine and Biology}, vol.~54, no.~23, 2009.

\bibitem{Asslander2021}
J.~Assl{\"{a}}nder, C.~Gultekin, S.~Flassbeck, S.~J. Glaser, and D.~K. Sodickson, ``{Generalized Bloch model: A theory for pulsed magnetization transfer},'' \emph{Magnetic Resonance in Medicine}, no. July, pp. 1--15, nov 2021.

\bibitem{Asslander2023}
J.~Assl{\"{a}}nder, A.~Mao, E.~S. Beck, F.~{La Rosa}, R.~W. Charlson, T.~M. Shepherd, and S.~Flassbeck, ``{On multi-path longitudinal spin relaxation in brain tissue},'' \emph{arXiv}, p. 2301.08394, 2023.

\bibitem{Mao2023}
A.~Mao, S.~Flassbeck, C.~Gultekin, and J.~Asslaender, ``{Cram{\'{e}}r-Rao Bound Optimized Linear Bases for Low-Rank Subspace Reconstruction},'' \emph{Proc. Intl. Soc. Mag. Reson. Med.}, 2023.

\bibitem{Vidal_2016}
R.~Vidal, Y.~Ma, and S.~Sastry, \emph{Generalized Principal Component Analysis}.\hskip 1em plus 0.5em minus 0.4em\relax Springer New York, 2016.

\bibitem{Li2019}
X.~Li, S.~Wang, and Y.~Cai, ``{Tutorial: Complexity analysis of Singular Value Decomposition and its variants},'' \emph{arXiv}, p. 1906.12085, jun 2019.

\bibitem{Chan2009}
R.~W. Chan, E.~A. Ramsay, C.~H. Cunningham, and D.~B. Plewes, ``Temporal stability of adaptive 3d radial {MRI} using multidimensional golden means,'' \emph{Magnetic Resonance in Medicine}, vol.~61, no.~2, pp. 354--363, Feb. 2009.

\bibitem{Flassbeck2021}
S.~Flassbeck and J.~Assl{\"{a}}nder, ``{Minimization of Eddy Current Related Artefacts in HSFP Sequences},'' \emph{arXiv}, p. 2203.06099, 2021.

\bibitem{Bojorquez2017}
J.~Z. Bojorquez, S.~Bricq, C.~Acquitter, F.~Brunotte, P.~M. Walker, and A.~Lalande, ``{What are normal relaxation times of tissues at 3 T?}'' \emph{Magnetic Resonance Imaging}, vol.~35, pp. 69--80, 2017.

\bibitem{Stanisz2005}
G.~J. Stanisz, E.~E. Odrobina, J.~Pun, M.~Escaravage, S.~J. Graham, M.~J. Bronskill, and R.~M. Henkelman, ``{T1, T2 relaxation and magnetization transfer in tissue at 3T},'' \emph{Magnetic Resonance in Medicine}, vol.~54, no.~3, pp. 507--512, 2005.

\bibitem{Ma2017}
D.~Ma, S.~Coppo, Y.~Chen, D.~F. McGivney, Y.~Jiang, S.~Pahwa, V.~Gulani, and M.~A. Griswold, ``{Slice profile and B1 corrections in 2D magnetic resonance fingerprinting},'' \emph{Magnetic Resonance in Medicine}, vol.~78, no.~5, pp. 1781--1789, 2017.

\bibitem{Malik2016}
S.~J. Malik, A.~Sbrizzi, H.~Hoogduin, and J.~V. Hajnal, ``{Equivalence of EPG and Isochromat-based simulation of MR signals},'' \emph{Proc. Intl. Soc. Mag. Reson. Med.}, p. 3196, 2016.

\bibitem{Marquardt1963}
\BIBentryALTinterwordspacing
D.~W. Marquardt, ``{An Algorithm for Least-Squares Estimation of Nonlinear Parameters},'' \emph{Journal of the Society for Industrial and Applied Mathematics}, vol.~11, no.~2, pp. 431--441, jun 1963. [Online]. Available: \url{http://epubs.siam.org/doi/10.1137/0111030}
\BIBentrySTDinterwordspacing

\bibitem{Knopp2021}
T.~Knopp and M.~Grosser, ``{MRIReco.jl: An MRI reconstruction framework written in Julia},'' \emph{Magnetic Resonance in Medicine}, vol.~86, no.~3, pp. 1633--1646, 2021.

\bibitem{Wajer2001}
F.~Wajer and K.~Pruessmann, ``{Major Speedup of Reconstruction for Sensitivity Encoding with Arbitrary Trajectories},'' \emph{Proc. Intl. Soc. Mag. Reson. Med.}, vol.~7, no.~2, 2001.

\bibitem{Baron2018}
C.~A. Baron, N.~Dwork, J.~M. Pauly, and D.~G. Nishimura, ``{Rapid compressed sensing reconstruction of 3D non‐Cartesian MRI},'' \emph{Magnetic Resonance in Medicine}, vol.~79, no.~5, pp. 2685--2692, may 2018.

\bibitem{Uecker2014}
M.~Uecker, P.~Lai, M.~J. Murphy, P.~Virtue, M.~Elad, J.~M. Pauly, S.~S. Vasanawala, and M.~Lustig, ``{ESPIRiT - An eigenvalue approach to autocalibrating parallel MRI: Where SENSE meets GRAPPA},'' \emph{Magnetic Resonance in Medicine}, vol.~71, no.~3, pp. 990--1001, 2014.

\bibitem{Trzasko2011}
J.~Trzasko and A.~Manduca, ``{Local versus Global Low-Rank Promotion in Dynamic MRI Series Reconstruction},'' \emph{Proc. Intl. Soc. Mag. Reson. Med.}, vol.~24, no.~7, p. 4371, 2011.

\bibitem{Zhang2015}
T.~Zhang, J.~M. Pauly, and I.~R. Levesque, ``{Accelerating parameter mapping with a locally low rank constraint},'' \emph{Magnetic Resonance in Medicine}, vol.~73, no.~2, pp. 655--661, 2015.

\bibitem{Beck2009}
A.~Beck and M.~Teboulle, ``{A Fast Iterative Shrinkage-Thresholding Algorithm},'' \emph{Society for Industrial and Applied Mathematics Journal on Imaging Sciences}, vol.~2, no.~1, pp. 183--202, 2009.

\bibitem{Rakshit2021}
S.~Rakshit, K.~Wang, and J.~I. Tamir, ``{A GPU-accelerated Extended Phase Graph Algorithm for differentiable optimization and learning},'' \emph{Proc. Intl. Soc. Mag. Reson. Med.}, 2021.

\bibitem{Cuomo2022}
S.~Cuomo, V.~S. di~Cola, F.~Giampaolo, G.~Rozza, M.~Raissi, and F.~Piccialli, ``{Scientific Machine Learning through Physics-Informed Neural Networks: Where we are and What's next},'' \emph{arXiv}, p. 2201.05624, 2022.

\bibitem{BART}
\BIBentryALTinterwordspacing
M.~Uecker, F.~Ong, J.~I. Tamir, D.~Bahri, P.~Virtue, J.~Y. Cheng, T.~Zhang, and M.~Lustig, ``Bart: Version 0.2.09,'' 2015. [Online]. Available: \url{https://zenodo.org/record/31907}
\BIBentrySTDinterwordspacing

\bibitem{Wang2022b}
N.~Wang, T.~Cao, F.~Han, Y.~Xie, X.~Zhong, S.~Ma, A.~Kwan, Z.~Fan, H.~Han, X.~Bi, M.~Noureddin, V.~Deshpande, A.~G. Christodoulou, and D.~Li, ``{Free-breathing multitasking multi-echo MRI for whole-liver water-specific T1, proton density fat fraction, and R2star quantification},'' \emph{Magnetic Resonance in Medicine}, vol.~87, no.~1, pp. 120--137, 2022.

\bibitem{Ma2022}
S.~Ma, N.~Wang, Y.~Xie, Z.~Fan, D.~Li, and A.~G. Christodoulou, ``{Motion-robust quantitative multiparametric brain MRI with motion-resolved MR multitasking},'' \emph{Magnetic Resonance in Medicine}, vol.~87, no.~1, pp. 102--119, 2022.

\bibitem{Bouhrara2018}
M.~Bouhrara and R.~G. Spencer, ``{Fisher information and Cram{\'{e}}r-Rao lower bound for experimental design in parallel imaging},'' \emph{Magnetic Resonance in Medicine}, vol.~79, no.~6, pp. 3249--3255, 2018.

\bibitem{Nehorai2014}
A.~Nehorai and G.~Tang, \emph{{Encyclopedia of Systems and Control}}.\hskip 1em plus 0.5em minus 0.4em\relax Springer London, 2014.

\bibitem{Fletcher2000}
R.~Fletcher, \emph{Practical Methods of Optimization}.\hskip 1em plus 0.5em minus 0.4em\relax John Wiley {\&} Sons, Ltd, May 2000.

\bibitem{McAulay1971}
R.~J. McAulay and E.~M. Hofstetter, ``{Barankin Bounds on Parameter Estimation},'' \emph{IEEE Transactions on Information Theory}, vol.~17, no.~6, pp. 669--676, nov 1971.

\bibitem{Yang2018}
M.~Yang, D.~Ma, Y.~Jiang, J.~Hamilton, N.~Seiberlich, M.~A. Griswold, and D.~McGivney, ``{Low rank approximation methods for MR fingerprinting with large scale dictionaries},'' \emph{Magnetic Resonance in Medicine}, vol.~79, no.~4, pp. 2392--2400, 2018.

\bibitem{Kay1993}
S.~M. Kay, \emph{\BIBforeignlanguage{en}{Fundamentals of Statistical Signal Processing, Volume {I}: Estimation Theory}}.\hskip 1em plus 0.5em minus 0.4em\relax Philadelphia, PA: Prentice Hall, 03 1993.

\bibitem{Newey1994}
W.~K. Newey and D.~McFadden, ``Chapter 36: Large sample estimation and hypothesis testing,'' in \emph{Handbook of Econometrics}.\hskip 1em plus 0.5em minus 0.4em\relax Elsevier, 1994, pp. 2111--2245.

\bibitem{Wu1981}
C.-F. Wu, ``{Asymptotic Theory of Nonlinear Least Squares Estimation},'' \emph{The Annals of Statistics}, vol.~9, no.~3, pp. 501--513, may 1981.

\bibitem{Asslander2022}
J.~Assl{\"{a}}nder, C.~Gultekin, A.~Mao, X.~Zhang, Q.~Duchemin, K.~Liu, R.~W. Charlson, T.~Shepherd, C.~Fernandez-Granda, and S.~Flassbeck, ``{Rapid quantitative magnetization transfer imaging: utilizing the hybrid state and the generalized Bloch model},'' \emph{arXiv}, p. 2207.08259, jul 2022.

\bibitem{Fessler2002}
J.~Fessler, ``{Mean and variance of implicitly defined biased estimators (such as penalized maximum likelihood): applications to tomography},'' \emph{IEEE Transactions on Image Processing}, vol.~5, no.~3, pp. 493--506, mar 1996.

\end{thebibliography}

\end{document}